\documentclass[sigconf, nonacm]{acmart}

\newcommand{\eat}[1]{}

\usepackage{latexsym}
\usepackage{amsfonts}
\usepackage{amsmath}
\usepackage{xcolor}
\usepackage{colortbl}
\usepackage{xspace}
\usepackage{subfigure}
\usepackage{paralist}
\usepackage{enumerate}
\usepackage[color,matrix,arrow,all]{xy}
\usepackage{comment}
\usepackage{booktabs}
\usepackage{balance}

\usepackage{hhline}
\usepackage{listings}
\usepackage{array}
\usepackage{float}
\usepackage[flushleft]{threeparttable}

\usepackage{makecell}
\usepackage{xparse}
\usepackage{wrapfig}

\usepackage{fontawesome5}

\usepackage{arydshln}

\usepackage{tikz}
\usetikzlibrary{matrix,positioning,arrows.meta,backgrounds}

\usepackage{epsfig}
\usepackage{multirow}
\usepackage{url}

\usepackage{multirow}

\usepackage{quoting}

\usepackage{listings}
\usepackage{framed}
\usepackage{xcolor}
\usepackage{changepage}
\colorlet{shadecolor}{gray!20}

\definecolor{shadecolor}{RGB}{220,220,220}

\definecolor{inputcolor}{RGB}{255,139,35}
\definecolor{outputcolor}{RGB}{120,212,252}
\definecolor{embedcolor}{RGB}{254,127,156}
\definecolor{maskcolor}{RGB}{122,128,255}
\definecolor{ecolor}{RGB}{58,149,54}

\definecolor{highcolor}{RGB}{255,153,153}
\definecolor{midcolor}{RGB}{255,204,204}
\definecolor{lowcolor}{RGB}{204,229,255}

\usepackage{tikz}
\usetikzlibrary{shapes,snakes}
\usetikzlibrary{calc}

\usepackage[export]{adjustbox}% http://ctan.org/pkg/adjustbox

\definecolor{green}{RGB}{0,128,0}

\definecolor{yellow}{RGB}{255,200,18}

\definecolor{myred}{HTML}{EFB3B2}     
\definecolor{myyellow}{HTML}{FFE6C4}   
\definecolor{mygreen}{HTML}{CCFCC4}

\newcommand{\stab}{\vspace{0.8ex}\noindent}

\newcommand{\bi}{\begin{itemize}}
\newcommand{\ei}{\end{itemize}}

\newcommand{\be}{\begin{enumerate}}
\newcommand{\ee}{\end{enumerate}}
\newcommand{\beqn}{\begin{eqnarray*}}
\newcommand{\eeqn}{\end{eqnarray*}}

\newcommand{\stitle}[1]{\stab\noindent{\bf #1}}

\newcommand{\wrt}{\emph{w.r.t.}\xspace}

\newcommand{\sys}{EUR\xspace}

\makeatletter
    \newcommand\figcaption{\def\@captype{figure}\caption}
    \newcommand\tabcaption{\def\@captype{table}\caption}
\makeatother

\NewDocumentCommand{\nan}{ mO{} }{\textcolor{blue}{\textsuperscript{\textit{Nan}}\textsf{\textbf{\small[#1]}}}}
\NewDocumentCommand{\haixun}{ mO{} }{\textcolor{blue}{\textsuperscript{\textit{Haixun}}\textsf{\textbf{\small[#1]}}}}

\NewDocumentCommand{\yang}{ mO{} }{\textcolor{green}{\textsuperscript{\textit{yang}}\textsf{\textbf{\small[#1]}}}}

\NewDocumentCommand{\zzx}{ mO{} }{\textcolor{yellow}{\textsuperscript{\textit{zzx}}\textsf{\textbf{\small[#1]}}}}

\usepackage{amsmath}
\usepackage{multirow}
\usepackage[T1]{fontenc}
\usepackage[utf8]{inputenc}

\usepackage{microtype}

\usepackage[ruled,vlined]{algorithm2e}
\usepackage{booktabs} 
\usepackage{siunitx} 
\usepackage[most]{tcolorbox}

\usepackage{algpseudocode}

\usepackage{enumitem}
\usepackage{hyperref}
\hypersetup{colorlinks=true, linkcolor=blue, urlcolor=blue, citecolor=blue}
\usepackage{microtype}

\usepackage{pifont}
\usetikzlibrary{shapes, arrows.meta,positioning,decorations.pathreplacing}

\tikzstyle{layer} = [rectangle, draw, rounded corners, text centered, text width=8cm, minimum height=0.8cm, fill=blue!10]
\tikzstyle{subblock} = [rectangle, draw, rounded corners, text centered, text width=5cm, minimum height=0.6cm, fill=green!10]
\tikzstyle{arrow} = [->, thick]

\definecolor{customgreen}{HTML}{CCFCC4} 
\definecolor{custompink}{HTML}{FFCAC9}

\tcbuselibrary{breakable}

\tcbset{
    mystyle/.style={
        colback=gray!5,        
        colframe=black!75,
        coltitle=white,
        fonttitle=\bfseries\small,
        title={#1},
        arc=0mm, 
        boxrule=0.5pt,
        left=3mm, right=3mm, top=2mm, bottom=2mm,
        after skip=1.2em,
        fontupper=\footnotesize
    }
}

\newtcolorbox{assumptionbox}[1]{
  enhanced,
  sharp corners,
  boxrule=0pt,
  colback=black!6,
  colframe=black!6,
  left=4mm,
  right=4mm,
  top=1mm,
  bottom=1mm,
  borderline west={2pt}{0pt}{red!75!black},
  before upper={
    \textbf{#1}
    \par\vspace{1mm}
  }
}

\newtcolorbox{titlebox}[1]{
  enhanced,
  sharp corners,
  boxrule=0pt,
  colback=white,
  colframe=white,
  left=1mm,
  right=1mm,
  top=1mm,
  bottom=1mm,
  borderline west={2pt}{0pt}{black},
  before upper={
    \textbf{#1}
    \par\vspace{-2mm}
  }  
}

\newtcolorbox{samplebox}[1]{
  enhanced,
  sharp corners,
  boxrule=0pt,
  colback=white,
  colframe=white,
  left=1mm,
  right=1mm,
  top=-2mm,
  bottom=-2mm,
  borderline west={2pt}{0pt}{black},
}

\newcommand{\showDOI}[1]{\unskip}

\newcommand\vldbdoi{XX.XX/XXX.XX}

\newcommand\vldbavailabilityurl{URL_TO_YOUR_ARTIFACTS}
\title{Back to the Future: The Evolving Universal Relation Model}
\subtitle{--- Query First, Database Emerges ---}

\author{Nan Tang}
\affiliation{
    \institution{HKUST (GZ), China}
    \city{}
    \country{}
}
\email{nantang@hkust-gz.edu.cn}

\author{Haixun Wang}
\affiliation{
    \institution{Evenup, USA}
    \city{}
    \country{}
}
\email{haixun@gmail.com}

\begin{document}
%!TEX root = ../main.tex
\begin{abstract}
\textit{Document-backed analytics need databases that can be built as queries arrive.}
Enterprises increasingly ask analytical queries over messy document collections---contracts, invoices, reports, emails, policies, and tickets---where the relevant structure and facts are often not known in advance.
RAG can retrieve and summarize supporting evidence, but it does not by itself produce reusable, joinable, governed database state. Schema-first construction offers the opposite strength: once a database is built, it supports joins, aggregation, constraints, indexes, optimization, and transactions. Yet for document analytics, fixing the schema and extracting all possible fields upfront is both expensive and brittle, because later queries and new documents often surface concepts that were not modeled before.
We propose the \emph{Evolving Universal Relation Model} (\textbf{\sys}), a query-first model for document-backed analytics. Users ask natural language queries as if the document corpus exposed one conceptual relation over application concepts. Internally, the DBMS turns each query into controlled database evolution: it updates the logical schema when new concepts are needed, extracts and integrates the facts required to answer the query, and decides what durable database state should remain for future queries.
Thus, in \sys, a query is not only evaluated; it also helps construct the database. Document-backed analytics becomes query-first universal access plus workload-driven database schema and instance evolution.
\end{abstract}
\maketitle

% %%% do not modify the following VLDB block %%
% %%% VLDB block start %%%
% \pagestyle{\vldbpagestyle}
% \begingroup\small\noindent\raggedright\textbf{PVLDB Reference Format:}\\
% \vldbauthors. \vldbtitle. PVLDB, \vldbvolume(\vldbissue): \vldbpages, \vldbyear.\\
% \href{https://doi.org/\vldbdoi}{doi:\vldbdoi}
% \endgroup
% \begingroup
% \renewcommand\thefootnote{}\footnote{\noindent
% This work is licensed under the Creative Commons BY-NC-ND 4.0 International License. Visit \url{https://creativecommons.org/licenses/by-nc-nd/4.0/} to view a copy of this license. For any use beyond those covered by this license, obtain permission by emailing \href{mailto:info@vldb.org}{info@vldb.org}. Copyright is held by the owner/author(s). Publication rights licensed to the VLDB Endowment. \\
% \raggedright Proceedings of the VLDB Endowment, Vol. \vldbvolume, No. \vldbissue\ %
% ISSN 2150-8097. \\
% \href{https://doi.org/\vldbdoi}{doi:\vldbdoi} \\
% }\addtocounter{footnote}{-1}\endgroup
% %%% VLDB block end %%%

% %%% do not modify the following VLDB block %%
% %%% VLDB block start %%%
% \ifdefempty{\vldbavailabilityurl}{}{
% \vspace{.3cm}
% \begingroup\small\noindent\raggedright\textbf{PVLDB Artifact Availability:}\\
% The source code, data, and/or other artifacts have been made available at \url{\vldbavailabilityurl}.
% \endgroup
% }
% %%% VLDB block end %%%

%!TEX root = ../main.tex
\section{Introduction}
\label{sec:introduction}

\textit{The next enterprise database starts as documents.} Traditional databases start with schemas; document-backed analytics start with queries. In compliance reporting, contract tracking, policy monitoring, and similar settings, users ask over contracts, invoices, reports, emails, policies, and tickets before the right schema is known. These queries are analytical, not merely conversational: they require joins, aggregations, comparisons, and trends. Yet the needed concepts and facts are scattered across messy documents and often become clear only as new queries arrive.

\textit{RAG only reads; analytics need databases.} Retrieval-augmented generation (RAG) makes document collections searchable and conversational, but it does not provide a database contract. It retrieves evidence into a prompt and produces a response; it does not, by itself, leave behind stable identifiers, typed tuples, relationships, constraints, versions, indexes, or governed provenance. Analytical queries should not repeatedly reconstruct the same facts from text snippets. They need reusable database that can be joined, aggregated, checked, maintained, and optimized.

\textit{Schema-first construction freezes the design too early.} The natural counter-move is to build the database upfront: declare entities and relationships, design tables, define mappings, and run extractors. This restores familiar DBMS benefits, but only if the schema, mappings, and required facts are already known. Classical models provide the right ideals, but under a schema-first assumption: Entity-Relationship (ER)~\cite{chen1976er} model captures entities, relationships, attributes, and constraints, while the relational model~\cite{codd1970relational} provides declarative access and data independence over a declared schema. What is missing is a contract that preserves these ideals while the database is still being discovered from documents and queries.

\textit{Research gap: database construction should become query-first.}
Document-backed analytics need a stable query interface over application concepts, while the DBMS is still deciding what tables exist, where attributes belong, which facts to extract, and what durable state to keep. The challenge is not better retrieval alone, nor better upfront extraction alone. It is query-driven database construction: each query may reveal new concepts, require new facts, and improve the reusable database state for future queries.

Pay-as-you-go dataspaces already challenged the need for complete
semantic integration before useful access
~\cite{franklin2005dataspaces,madhavan2007paygo}.
The question we target is therefore not whether construction can
be incremental, but how a DBMS should govern it: when a query
requires concepts or facts that are not yet represented, how
should the query-facing vocabulary, logical schema, and
materialized instance evolve together while preserving logical
path independence? Document-backed analytics needs a contract
that makes these changes controlled effects of query processing,
so that answering a query also builds reusable database state.

\textit{Back to the future: The \sys model.} We propose the \textit{Evolving Universal Relation Model} (\textbf{\sys}, pronounced ``your''): a query-first revival of the universal relation model (URM)~\cite{ullman1982ur} for document-backed analytics. Users pose queries as if the document collection exposed a single evolving universal relation over application concepts. Internally, the DBMS maintains three coupled levels: an evolving universal relation that defines the user-facing query interface; a logical schema that organizes concepts into entity and relationship tables; and a materialized instance containing extracted, integrated facts. Each query both returns an answer and triggers a controlled database evolution. Unlike schema-on-read, this interpretation is retained as durable database state; unlike schema-first ETL, the design evolves
as the workload reveals what is needed.

% \textit{Outline.}
% Section~\ref{sec:example} motivates the problem.
% Section~\ref{sec:model} defines \sys.
% Section~\ref{sec:evolution} presents query-driven evolution.
% Sections~\ref{sec:discussion}--\ref{sec:research-agenda} discuss database principles and system support.
% Section~\ref{sec:call-to-arms} concludes.
%!TEX root = ../main.tex
\section{A Motivating Example}
\label{sec:example}

Next, we illustrate \sys with a small course-analytics example. The key point is that a query does not directly create tables. It first reveals application-level vocabulary: \textit{attribute roles} and \textit{semantic connections}. \sys merges these revealed roles and connections into the \textit{evolving universal relation}. The DBMS then derives a logical schema by decomposing this relation according to the semantic connections and (available) functional dependencies. Finally, the system extracts the document-backed facts needed for the current queries and materializes them as reusable database state.

Consider an example about course analytics over a document collection $F$. Suppose $F$ contains the following document snippet.

%\vspace{-2ex}
\begin{center}
\noindent\fbox{%
\parbox{\columnwidth}{\small
\textbf{Text snippet:}
In 2026, the CS department in the School of Engineering offers \textsf{CS101}, titled \emph{Introduction to AI}, and \textsf{CS201}, titled \emph{Database Systems}. \textsf{CS101} has 3 credits, and \textsf{CS201} has 4 credits. Alice Sara is a CS major and takes both \textsf{CS101} and \textsf{CS201}.
}%
}
\end{center}

A workload with two queries arrive over time:

%\vspace{-2ex}
\begin{center}
\noindent\fbox{%
\parbox{\columnwidth}{\small
$Q_1$: Which CS courses are offered in 2026 by the School of Engineering?\\
$Q_2$: Which of these courses does Alice Sara, a CS major, take, and how\\
    many credits is each course?
}%
}
\end{center}

For simplicity, let's assume that initially, the database is empty.

\subsection{First Query: Revealing the Course World}

Query $Q_1$ asks for courses, but it also mentions the conditions under which those courses should be selected: department, school, and year. The system therefore begins by identifying the attribute roles and semantic connections needed to interpret the query.

\stitle{Attribute roles and semantic connections.}
The query reveals five \textit{attribute roles}:
\begin{samplebox}{}
{\small
\[
\begin{array}{rcl}
C(\text{course})     &:& \mathsf{CourseCode}\\
T(\text{title})      &:& \mathsf{CourseTitle}\\
D(\text{department}) &:& \mathsf{DepartmentName}\\
S(\text{school})     &:& \mathsf{SchoolName}\\
Y(\text{year})       &:& \mathsf{OfferingYear}
\end{array}
\]
}
\end{samplebox}

The query also reveals semantic connections among these roles, represented as \textit{predicates} $P$ over subsets of the attribute roles:

\begin{samplebox}{Attribute roles revealed and their semantic connections}
{\small
\[
\begin{array}{rcll}
P_{\mathsf{course}}(C,T)
&:& \text{course } C \text{ has title } T,\\[2pt]
P_{\mathsf{department}}(D,S)
&:& \text{department } D \text{ belongs to school } S,\\[2pt]
P_{\mathsf{offering}}(D,C,Y)
&:& \text{department } D \text{ offers course } C \text{ in year } Y.
\end{array}
\]
}
\end{samplebox}

\stitle{Universal relation.}
The revealed roles and semantic connections are merged into the first query-facing \textit{universal relation} $U_1$ \wrt $Q_1$~\cite{ullman1982ur}. $U_1$ is a conceptual relation, not a stored wide table:

\begin{samplebox}{Universal relation}
{\small
\[
\begin{array}{l}
U_1 (C,T,D,S,Y). \\
U_1 = \{(c,t,d,s,y) \mid P_{\mathsf{course}}(c,t) 
\!\wedge\! P_{\mathsf{department}}(d,s)
\!\wedge\! P_{\mathsf{offering}}(d,c,y)
\}.
\end{array}
\]
}
\end{samplebox}

It lets the query refer to $(C,T,D,S,Y)$ without exposing how the DBMS will later organize these roles into tables. Under $U_1$, the query is interpreted below, which is independent of any particular table decomposition:

\begin{samplebox}{Universal relation}
{\small
\[
\begin{array}{ll}
Q_1(U_1) =& \{(c,t) \mid (c,t,d,s,y)\in U_1 \wedge d=\textsf{``CS''} \\
         & \wedge s=\textsf{``Engineering''} \wedge y=2026\}.
\end{array}
\]
}
\end{samplebox}

The query has revealed the application vocabulary needed to answer it. At this stage, the system knows \textit{what} concepts exist, but not yet \textit{how} they should be organized into relational tables.

\stitle{Logical schema.}
After $U_1$ is established, the DBMS chooses a logical schema underneath it. The design is not obtained by creating one table per attribute role. Instead, the DBMS decomposes the universal relation according to the semantic connections and (available) functional dependencies as:

\begin{samplebox}{Logical schema design}
{\small
\[
\begin{array}{rl}
E_1
  &=
  \{
  \textbf{Course}(\textsf{CourseCode},\textsf{CourseTitle}), \\
  &\hspace{3ex}\textbf{Department}(\textsf{DepartmentName},\textsf{SchoolName})
  \},\\[2pt]
R_1
  &=
  \{
  \textbf{Offers}(\textsf{DepartmentName},\textsf{CourseCode},\textsf{OfferingYear})
  \}.
\end{array}
\]
}
\end{samplebox}

Here, \textbf{Course} and \textbf{Department} are entity tables, while \textbf{Offers} is a relationship table. The attribute \textsf{OfferingYear} is placed in \textbf{Offers} because it describes a department-course offering.

This separation is why \sys keeps the universal relation distinct from the logical schema. \(U_1\) is the \textit{user-facing} query interface: it presents concepts and predicates without exposing tables or join paths. \(S_1=(E_1,R_1)\) is the \textit{DBMS-facing} design: it decomposes those concepts into entity and relationship tables for better management.

\begin{figure}[t!]
   \centering
    \includegraphics[width=\columnwidth]{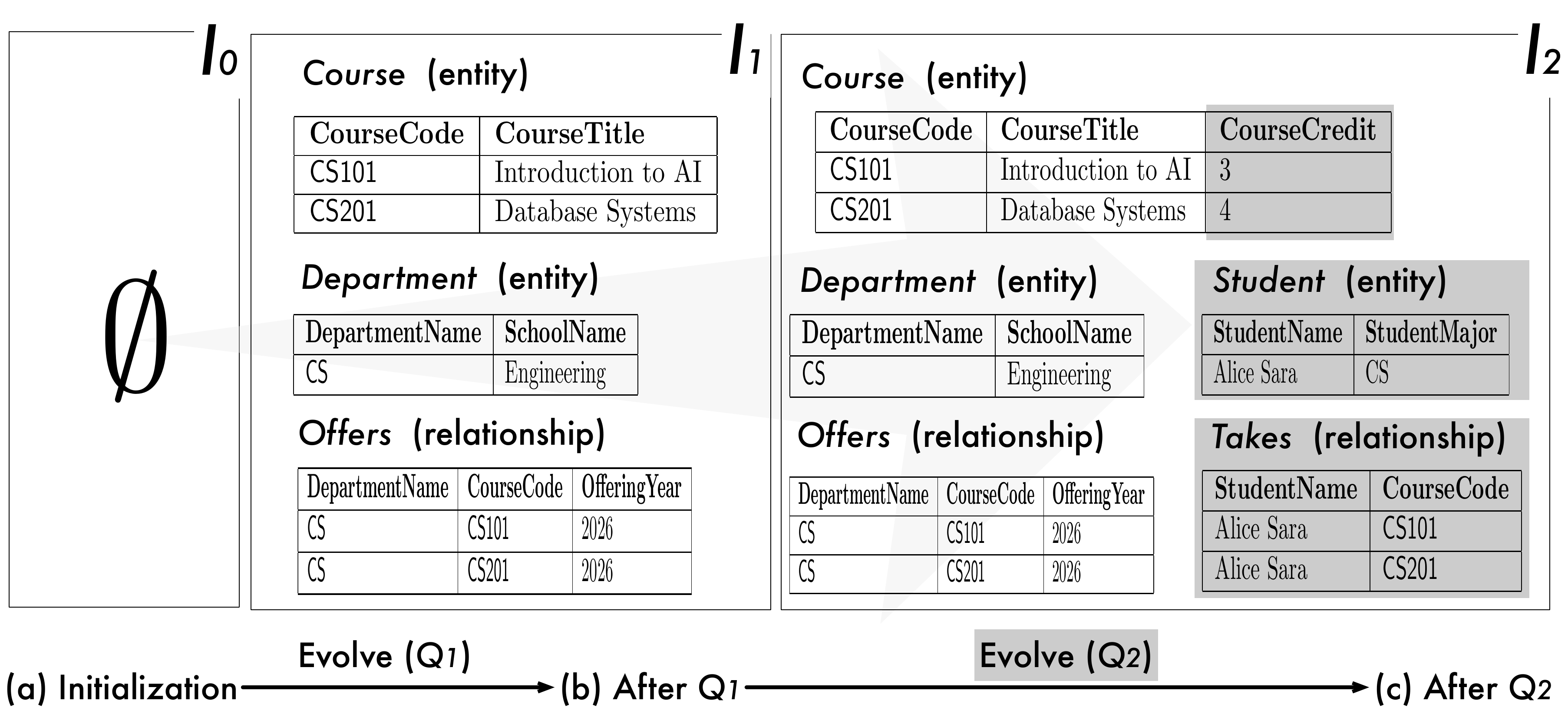}
    %\vspace{-2em}
    \caption{Sample database evolution guided by queries.}
    \label{fig:evolve}
    %\vspace{-1em}
\end{figure}

\stitle{Materialized instance.}
The logical schema \(S_1\) specifies what the DBMS can store, but the database is still empty. The system therefore identifies the materialization gap: the facts required to answer \(Q_1\) that are missing from the initial database instance \(I_0 = \emptyset\). It then extracts and integrates those facts from the document snippet, producing the materialized instance \(I_1\), as shown in Figure~\ref{fig:evolve}(b).

\subsection{Second Query: Reusing and Extending State}

Query $Q_2$ asks which of the courses identified by $Q_1$ Alice Sara takes, and how many credits each course has. The query therefore reuses part of the database produced by $Q_1$: the course role $C$ and the course-offering scope already materialized in $I_1$. At the same time, it reveals new information about students, enrollments, and course credits. As before, the system first interprets the query as attribute roles and semantic connections.

\stitle{\textit{New} attribute roles and semantic connections.}
The query $Q_2$ reuses course role $C$ from $U_1$ and reveals three new attribute roles:
\begin{samplebox}{}
{\small
\[
\begin{array}{rcl}
N(\text{student}) &:& \mathsf{StudentName}\\
M(\text{major})   &:& \mathsf{StudentMajor}\\
K(\text{credits}) &:& \mathsf{CourseCredits}
\end{array}
\]
}
\end{samplebox}

The new roles $N$, $M$, and $K$ are additional attributes over which the second query is interpreted.

Query $Q_2$ also reveals new semantic connections as predicates:
\begin{samplebox}{Attribute roles revealed and their semantic connections}
{\small
\[
\begin{array}{rcll}
P_{\mathsf{student}}(N,M)
&:& \text{student } N \text{ has major } M,\\[2pt]
P_{\mathsf{takes}}(N,C)
&:& \text{student } N \text{ takes course } C,\\[2pt]
P_{\mathsf{credits}}(C,K)
&:& \text{course } C \text{ has } K \text{ credits.}
\end{array}
\]
}
\end{samplebox}

Thus, $Q_2$ extends the application vocabulary of $Q_1$ from courses and offerings to students, enrollments, and credits.

\stitle{\textit{Evolving} universal relation.}
The newly revealed roles and predicates are merged with the existing universal relation. The resulting universal relation $U_2$ \wrt $(Q_1, Q_2)$ extends $U_1(C,T,D,S,Y)$ as:

\begin{samplebox}{Universal relation}
{\small
\[
\begin{array}{l}
U_2(C,T,D,S,Y,N,M,K). \\
U_2=\{(c,t,d,s,y,n,m,k)\mid
P_{\mathsf{course}}(c,t)
\wedge
P_{\mathsf{department}}(d,s) \\
\wedge
P_{\mathsf{offering}}(d,c,y)
\wedge\
P_{\mathsf{student}}(n,m)
\wedge
P_{\mathsf{takes}}(n,c)
\wedge
P_{\mathsf{credits}}(c,k)
\}.
\end{array}
\]
}
\end{samplebox}

The universal relation is again the query-facing interpretation of the workload. The user asks over the attributes $N,M,C,K$ together with the previously revealed course-offering attributes. The DBMS hides whether those roles are implemented by extending an existing table, creating a new entity table, or adding a relationship table.

\stitle{\textit{Evolving} logical schema.}
After $U_2$ is established, the DBMS extends the logical schema underneath it. The design reuses the entity and relationship tables from $S_1$ and adds only what is needed for the new query, annotated by $\overline{X}$.

\begin{samplebox}{Logical schema design}
{\small
\[
\begin{array}{rl}
E_2
  &=
  \{
  \textbf{Course}(\textsf{cname},\textsf{title},\overline{\textsf{credits}}),
  \textbf{Department}(\textsf{dname},\textsf{school}),\\
  &\qquad
  \overline{\textbf{Student}(\textsf{sname},\textsf{major})}
  \},\\[2pt]
R_2
  &=
  \{
  \textbf{Offers}(\textsf{dname},\textsf{cname},\textsf{year}),
  \overline{\textbf{Takes}(\textsf{sname},\textsf{cname})}
  \}.
\end{array}
\]
}
\end{samplebox}

This shows how schema evolution is driven by the query but mediated through the universal relation. $U_2$ manages the expanded query-facing attributes and predicates, while $S_2 =(E_2, R_2)$ is one DBMS-chosen decomposition into entity and relationship tables.

\stitle{\textit{Evolving} materialized instance.}
The updated logical schema $S_2=(E_2,R_2)$ says what can be stored, but not all needed facts have been materialized yet. The database can reuse the course, department, and offering facts already extracted for $Q_1$. It only needs to extract the facts newly required by $Q_2$: course credits, student record, and the enrollment facts. Given the above requirement, the system then extracts required facts as $I_2$ shown in Figure~\ref{fig:evolve}(c).
%\input{sections/sec_example_legal}
%!TEX root = ../main.tex
% \begin{assumptionbox}{A New Contract for Query-Driven Databases}
% {
% \textit{Ask over concepts; return an answer; evolve the database.}
% }
% \end{assumptionbox}
\section{The \sys Model}
\label{sec:model}

Classical Universal Relation Model (URM) assumes that the universal relation is defined upfront by its predicates and dependencies~\cite{ullman1982ur}. 
Our proposed \textit{Evolving Univeral Model} (\sys) keeps the one-relation interface, but changes how that relation is built: instead of requiring the full design before querying, the database evolves as queries arrive over the document collection. We now make this precise.

\stitle{Evolution step.}
The \sys model captures application-bounded, query-driven evolution for document-backed databases. Given a coherent application $\mathcal{A}$, a document collection $F$\footnote{For simplicity, we first treat $F$ as fixed. In practice, \sys also allows document changes, denoted $\Delta F$, which trigger new document-to-database extraction steps.}, and an evolving workload $(Q_1,\ldots,Q_n)$, \sys treats each workload increment $Q_i$ as both a query to answer and a controlled database evolution step:
\[
D_{i-1} \overset{Q_i,F}{\leadsto} D_i .
\]
The application $\mathcal{A}$ is the boundary of evolution: it defines the recurring intents, answer forms, governance rules, and evidence
standards that make change controlled rather than open-ended. A workload increment $Q_i$ may be a natural-language query, a batch of related queries, or an emerging query pattern. It is interpreted over application concepts, not over the tables that happen to exist.

\stitle{Database state.}
After processing $Q_1, \ldots, Q_i$, the database state is:
\[
D_i=(U_i,S_i,I_i).
\]
Here, $U_i$ is the evolving universal relation: the user-facing conceptual relation. $S_i=(E_i,R_i)$ is the evolving logical design, organized into entity and relationship schemas. $I_i$ is the materialized instance extracted and integrated from documents.

\begin{figure}[t!]
%\vspace{-1em}
\centering
\includegraphics[width=\columnwidth]{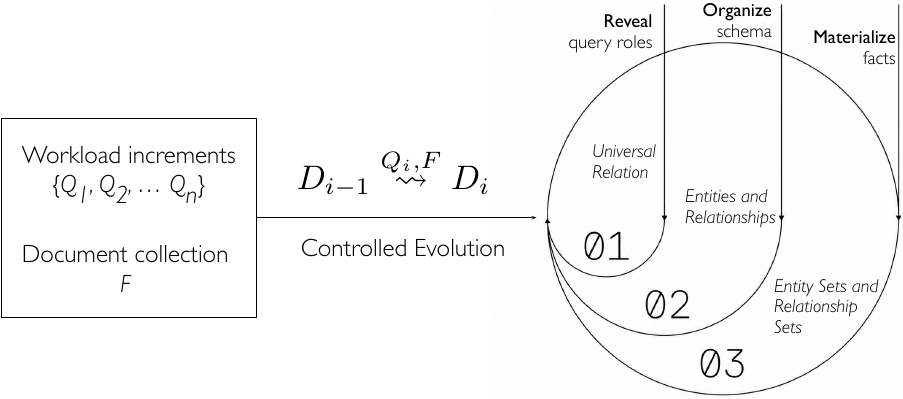}
%\vspace{-2.5em}
\caption{Query-driven database evolution.}
\label{fig:overview}
%\vspace{-1em}
\end{figure}

\stitle{Three coupled levels.}
As shown in Figure~\ref{fig:overview}, each transition $D_{i-1}\overset{Q_i,F}{\leadsto}D_i$ is implemented through three coupled levels:
(1) \textbf{\textit{Reveal:}} expose new attribute roles, concepts, predicates, and semantic connections, thereby evolving $U_{i-1}$ to $U_i$;
(2) \textbf{\textit{Organize:}} place the revealed concepts into entity and relationship schemas, thereby evolving $S_{i-1}$ to $S_i$; and
(3) \textbf{\textit{Materialize:}} extract, reconcile, validate, and store document-backed facts, thereby evolving $I_{i-1}$ to $I_i$.

Thus, users continue to ask over a conceptual universal relation, while the DBMS evolves the universal relation, logical design, and materialized instance underneath.

\stitle{Answer plus durable evolution.}
Each step has two outcomes: an answer to the current query and an updated durable database state $D_i$. Existing Text-to-SQL~\cite{nl2sql} techniques can help produce the answer once the needed schema and facts are available. The distinctive role of \sys is to decide what state the query should induce when the needed roles, schemas, or facts are missing.

\stitle{Latent limit state.}
For application $\mathcal{A}$, document collection $F$, and long-running workload $W_\infty=(Q_1,Q_2,\ldots)$, the transition sequence
\[
D_0 \overset{Q_1,F}{\leadsto} D_1
\overset{Q_2,F}{\leadsto} \cdots \leadsto D_\infty
\]
can be viewed as revealing finite approximations to an eventual application-bounded latent state $D_\infty=(U_\infty,S_\infty,I_\infty)$. Here, $U_\infty$ denotes the latent universal relation: the semantic closure of the attribute roles and connections that the workload would eventually make relevant under $\mathcal{A}$. Each finite state $D_i$ is the governed approximation reached after the prefix $(Q_1,\ldots,Q_i)$, and each query decides which part of the latent state is worth revealing, organizing, and materializing now.

\stitle{Starting assumptions.}
An \sys database may start from either an initial database or an empty database. In production, applications often provide partial schemas, mappings,
extracted facts, or coverage metadata upfront; this initializes some of $U_0$, $S_0$, and $I_0$.

%!TEX root = ../main.tex
\section{Reveal, Organize, Materialize}
\label{sec:evolution}

As shown in Figure below, each query is handled as a controlled evolution step with three parts. First, \sys reveals the concepts needed to understand the query. Second, it organizes those concepts into entity and relationship schemas. Third, it materializes the missing document-backed facts needed to answer the query.

\begin{figure}[h!]
%\vspace{-1em}
\centering
\resizebox{\columnwidth}{!}{%
\begin{tikzpicture}[
    >=Latex,
    font=\small,
    box/.style={
        draw,
        rounded corners=1pt,
        thick,
        inner sep=2pt,
        align=center,
        fill=gray!5,
        text width=3.8cm
    }
]

\node[box] (u) {
    \textbf{Reveal}\\[1mm]
    $\Delta U_i=\textsc{Reveal}(Q_i,U_{i-1})$\\[1mm]
    $U_{i-1}\overset{\Delta U_i}{\leadsto}U_i$
};

\node[box, right=5mm of u] (s) {
    \textbf{Organize}\\[1mm]
    $\Delta S_i=\textsc{Organize}(S_{i-1},U_i,\Delta U_i)$\\[1mm]
    $S_{i-1}\overset{\Delta S_i}{\leadsto}S_i$
};

\node[box, right=5mm of s] (i) {
    \textbf{Materialize}\\[1mm]
    $\Delta I_i=\textsc{Extract}(F,\Omega_i)$\\[1mm]
    $I_{i-1}\overset{\Delta I_i}{\leadsto}I_i$
};

\draw[->, thick] (u.east) -- (s.west);
\draw[->, thick] (s.east) -- (i.west);

\end{tikzpicture}%
}
\vspace{-2em}
\end{figure}

\subsection{Reveal: The Universal Relation Grows}
\label{subsec:evolving_relation}

At step $i$, $U_i$ is the conceptual relation exposed to users. It records the query-facing vocabulary revealed so far: attribute roles that can be returned, filtered, grouped, or aggregated, and semantic connections that explain how those roles belong together. It is not a stored wide table and not the logical schema.

\stitle{Query-revealed roles and connections.}
When $Q_i$ arrives, \sys identifies additional roles and connections needed to interpret it:
\[
  \Delta U_i=\textsc{Reveal}(Q_i,U_{i-1}).
\]
The delta may introduce missing roles and connections, or reuse and refine concepts already present in $U_{i-1}$.

\stitle{Universal relation transition.}
The revealed delta is semantically merged into the accumulated universal relation:
\[
  U_i=\textsc{Merge}(U_{i-1},\Delta U_i)
  \qquad \implies \qquad
  U_{i-1}\overset{\Delta U_i}{\leadsto}U_i .
\]
The merge resolves aliases, reused concepts, compatible predicates, refined meanings, and provenance-related roles under the application interpretation policy. The result $U_i$ is the updated query-facing contract; storage decisions are deferred to the logical-schema step.

\subsection{Organize: Roles Become Relations}
\label{subsec:evolving_entity}

The universal relation is the application-facing query contract. The logical schema is the DBMS-facing organization of that contract. As $U_i$ grows, \sys places revealed roles and semantic connections into entity and relationship schemas.

\stitle{Incremental normalization.}
In traditional databases, normalization decomposes a known wide relation into smaller tables around entities and relationships. In \sys, this organization is incremental, because the universal relation and the evidence behind it arrive over time. Semantic connections in $U_i$, together with available functional dependencies, guide whether a revealed role becomes an entity attribute, a relationship attribute, a key/foreign-key role, or part of a new relationship schema.

Given $S_{i-1}$, the updated universal relation $U_i$, and the newly revealed delta $\Delta U_i$, \sys computes the schema delta:
\[
  \Delta S_i=(\Delta E_i,\Delta R_i)
  =
  \textsc{Organize}(S_{i-1},U_i,\Delta U_i).
\]
It then applies this delta to obtain the updated logical schema:
\[
  S_i=(E_i,R_i)
  =
  \textsc{Merge}(S_{i-1},\Delta S_i)
  \qquad \implies \qquad
  S_{i-1}\overset{\Delta S_i}{\leadsto}S_i .
\]
Here, $E_i$ is the set of entity-table schemas and $R_i$ is the set of
relationship-table schemas after step $i$. In the monotonic version
considered here, \sys only extends the schema by introducing entities,
attributes, or relationships. More general evolution may revise earlier
choices, such as moving attributes across tables or splitting and merging
schemas; such changes require schema mappings and instance migration across
versions.

%Here, $E_i$ is the set of entity-table schemas and $R_i$ is the set of relationship-table schemas after step $i$. In the monotonic version considered here, \sys may introduce entity schemas, add attributes to existing entity schemas, or introduce relationship schemas, but it does not revise earlier normalization choices\footnote{More general forms of \sys may support non-monotonic evolution: moving an attribute from an entity to a relationship, splitting or merging entity schemas, or promoting a relationship into an entity. Such changes require schema mappings and instance migration across versions.}.

\subsection{Materialize: Gaps Become Facts}
\label{subsec:extraction}

After $U_i$ determines what users can ask and $S_i=(E_i,R_i)$ determines how those roles and connections are organized, the instance $I_i$ must contain enough document-backed facts to support $Q_i$. Materialization is the execution side of evolution: the query exposes what is missing, and the system turns that gap into extracted, integrated, and reusable state.

\stitle{Materialization gaps.}
Given the evolved schema and the previous instance, \sys derives a materialization gap $\Omega_i$ and passes it to the document-to-database extraction layer:
\[
\begin{array}{rl}
\Omega_i
  &= \textsc{MaterializationGap}(Q_i, \Delta S_i, I_{i-1}),\\
\Delta I_i
  &= \textsc{Extract}(F,\Omega_i),\\
I_i
  &= \textsc{Merge}(I_{i-1},\Delta I_i)
     \qquad \implies \qquad
     I_{i-1}\overset{\Delta I_i}{\leadsto}I_i .
\end{array}
\]
Here, $\Omega_i$ specifies the missing role values, entity tuples, relationship tuples, predicate regions, or evidence obligations needed to answer $Q_i$ under the evolved universal relation and schema. The instance delta $\Delta I_i$ contains newly extracted and reconciled facts over $S_i$, together with the evidence and provenance metadata needed to justify them.

\stitle{Materialization boundary and policy.}
\sys decides what the query reveals and how it changes $U_i$ and $S_i$; the document-to-database layer decides how to close the resulting gap in $I_i$. DataMosaic~\cite{datamosaic2026} studies how new queries and document updates trigger extraction, reconciliation, validation, and integration of facts.

Two materialization policies bracket the cost-reuse tradeoff. A \textbf{\textit{schema-complete}} policy extracts all facts for newly revealed schema fragments, maximizing future reuse but risking over-extraction. A \textbf{\textit{query-guided}} policy extracts only what is needed to answer $Q_i$ while leaving behind useful reusable facts.
Detailed extraction planning is left to the document-to-database layer.
%!TEX root = ../main.tex
\section{Database Principles, Rewired}
\label{sec:discussion}

\textit{\sys does not discard classical database principles; it changes when they must operate.}
Documents provide the evidence; workloads provide the design signal. A DBMS must connect the two while preserving the principles that still matter: identity, logical path independence, physical data independence, optimization, provenance, constraints, and transactions. In \sys, the contract changes from managing a mostly fixed database to managing the transition $D_{i-1} \overset{Q_i,F}{\leadsto} D_i$ as part of query processing. The universal relation, the logical entity/relationship design, the materialized instance, and the coverage state must evolve with the workload, while users continue to ask over a stable conceptual interface.

\subsection{What Databases Already Got Right}

Database systems already provide much of what document-backed analytics lacks. Chen's ER model gives a language for stable entities, relationships, attributes, and constraints~\cite{chen1976er}. Codd's relational model separates logical access from physical representation and access paths~\cite{codd1970relational}. Ullman's Universal Relation Model (URM) adds the user-facing ideal of logical path independence: users query attributes and predicates without choosing stored relations or join paths~\cite{ullman1982ur}. Together, these ideas express the contract that \sys inherits: users should ask about application concepts, while the DBMS manages logical design and physical access underneath.

This separation is exactly what document-backed applications need, but under a different timing assumption. Recent work argues that DBMSs should elevate ER concepts as first-class abstractions~\cite{deshpande2025beyond}, which aligns with \sys: document analytics needs stable things, relationships, identifiers, constraints, and provenance, not merely text chunks, prompt contexts, or flat extracted tables. Relational systems also provide the execution discipline needed once state exists. Cost-based optimization~\cite{selinger1979access}, query answering using views and data integration~\cite{halevy2001answering}, and provenance management~\cite{buneman2001why} are all directly relevant to \sys. Text-to-SQL~\cite{nl2sql} techniques can help answer the current query once the needed schema and facts are available. The distinctive problem in \sys is what happens when they are not available: the query must also induce universal-relation fragments, logical schema deltas, materialization gaps, extraction tasks, and durable state for future use.

\stitle{Relation to pay-as-you-go.}
Dataspaces and PAYGO established that useful access need not wait
for complete semantic integration, and that integration effort
should be invested incrementally as needs emerge
~\cite{franklin2005dataspaces,madhavan2007paygo}.
EUR shares this philosophy but targets a specific database-model
question. Whereas PAYGO addresses web-scale heterogeneity through
schema clusters, approximate mappings, and query routing, EUR
targets application-bounded document analytics through an evolving
universal relation. Each workload increment may jointly extend
the query-facing vocabulary, reorganize its logical representation,
and materialize missing facts. These changes are coupled effects
of query processing, not independent improvements to source
integration. EUR thus specializes the pay-as-you-go philosophy
into a model and systems agenda for governing the database state
that queries leave behind.

\subsection{Why Schema-First Breaks}

Classical DBMS contracts are schema-first. A catalog records relations after they are declared; an optimizer chooses plans over existing tables and indexes; storage assumes known layouts; maintenance begins from base data already inside the system. \sys changes this order. A workload increment may reveal a new attribute, add an entity type, extend an existing entity, introduce a relationship table, or require facts that have not yet been extracted from documents. The user should not have to follow these changes. Logical path independence means that the user continues to ask over the conceptual application domain, while the DBMS decides how concepts map to evolving tables, joins, stored facts, and coverage state.

The individual actions are not exotic. SQL already supports table creation, schema alteration, tuple insertion, and merge. \sys therefore does not require a new storage substrate merely to represent evolving tables. The challenge is control: today these operations are treated as occasional administrative changes or ordinary transactional writes, not as planned effects of query processing. Current DBMSs are not optimized for workloads in which queries repeatedly trigger small schema changes, document extraction jobs, new relationship tables, coverage updates, and new physical access paths.

Thus familiar components must be reinterpreted. The catalog must become an evolution catalog: it must record not only schemas, but also universal-relation roles, aliases, workload-induced scopes, schema versions, document mappings, materialization coverage, provenance, and confidence. Storage engines must support schemas whose attributes and relationships are discovered gradually. Record layouts, column files, compression dictionaries, statistics, and metadata generally assume that columns are declared first and changed infrequently. In \sys, adding attributes, creating relationship tables, backfilling columns, and reorganizing stored state become normal execution events rather than rare administrative operations.

Indexing exposes the same mismatch. Classical indexes such as B$^+$-trees are excellent for range predicates over stable attributes, but they are not designed for attributes, tuples, and relationships discovered on demand. Creating, updating, and discarding indexes over newly extracted columns may cost more than the query they support.
The optimizer must therefore choose among reuse, natural-language-to-query compilation over existing state, universal-relation evolution, logical schema evolution, extraction, validation, indexing, and over-materialization. Its cost model must account not only for disk I/O and CPU, but also for extraction cost, uncertainty, coverage gaps, and expected future reuse.

\subsection{Evolution as the New DBMS Primitive}

In \sys, a query should not merely consume an existing database; it should help shape the database that ought to exist. Each query exposes a small part of the application world, and the DBMS turns that exposure into lasting conceptual, logical, and materialized state. The user continues to ask over stable application concepts, while the system quietly lets the universal relation, schema, facts, coverage, and access structures grow with the workload.

This vision also repositions document extraction. The document-to-database layer is not an offline ETL stage, but the mechanism by which missing evidence becomes governed database state~\cite{datamosaic2026}. A query is complete only when it returns an answer and leaves behind reusable, validated, and accountable state for future queries.
%!TEX root = ../main.tex
\section{The Stack That Evolves}
\label{sec:research-agenda}

We now turn the model contract into a systems agenda: what DBMS stack should manage when the universal relation, logical schema, and materialized instance are all evolving?

\stitle{Storage for sparse, evolving coverage.}
Logical path independence is useful only if the physical layer can absorb evolution without exposing it to users. Row-oriented storage is a poor fit when attributes are added frequently and coverage is partial: adding an attribute may require changing layouts, rewriting records, refreshing statistics, and distinguishing true nulls from values not yet extracted. Columnar storage is a more natural starting point: newly revealed attributes can be appended as columns, missing values can remain sparse, and queries can scan only the attributes in scope. The columnar designs of C-Store~\cite{stonebraker2005cstore} and MonetDB~\cite{DBLP:journals/cacm/BonczKM08} provide useful foundations. The open problem is to adapt column storage to \emph{evolving coverage}: columns may be materialized only for selected entities, documents, or workload scopes, and backfilled later with provenance links and coverage metadata.

\stitle{Indexes that follow the workload.}
Hash indexes remain useful for point lookups and key-based joins, and B$^+$-trees remain attractive for range predicates over stable attributes. But \sys creates a different regime: attributes, tuples, and relationships are discovered on demand, and a newly extracted column may be useful for only a few queries. Building, updating, and discarding indexes may cost more than the query they support. Packed-memory-array layouts, as used in cache-oblivious indexing~\cite{bender2005cache}, reduce reorganization cost under insertions, while database cracking~\cite{idreos2007database}
incrementally reorganizes columns as queries arrive. \sys needs adaptive indexes that track workload-induced scopes, partial coverage, universal-relation predicates, and expected reuse as part of the evolution plan.

\stitle{Optimization as evolution planning.}
The optimizer cannot be only an access-path selector over fixed base tables. It must choose among answering over existing state, compiling natural language to queries, extending the universal relation, evolving the logical schema, extracting missing facts, building indexes, and over-materializing for reuse. Optimization therefore becomes a joint problem of query planning and evolution planning: the plan decides not only how to compute the answer, but also what durable state should remain afterward. The cost model must include document retrieval, LLM calls, extraction
failures, validation, entity resolution, index maintenance, schema migration, coverage updates, and expected reuse, in addition to I/O and CPU. The objective is no longer cheapest current execution; it is cheapest useful evolution.

\stitle{Normalization when meanings arrive late.}
%\sys reopens design questions that traditional database design settles upfront. 
In classical design, many-to-many relationships normally become relationship tables, while one-to-one and many-to-one relationships are often represented as attributes or foreign keys. In \sys, the cardinality and meaning of a relationship may not be known when it first appears in the universal relation. A system may initially materialize relationships uniformly as relationship tables; later workloads may reveal that some should collapse into entity attributes or foreign keys, while others should become richer relationship entities. The research problem is safe normalization and denormalization over partially materialized data while preserving provenance, coverage, schema versions, and previously answered-query semantics.

\stitle{Continuous cleaning and constraints.}
Data quality problems are amplified when facts are extracted incrementally from heterogeneous evidence. Extracted tuples may be duplicated, incomplete, contradictory, or expressed in incompatible units and formats. Normalization, transformation, entity resolution, and constraint checking therefore become continuous operations rather than one-time cleaning steps. Constraints must
distinguish database violations from evidence uncertainty: a key conflict may indicate a duplicate, an extraction error, an ambiguous entity, or genuinely conflicting evidence. \sys needs constraints that govern both schema evolution and materialized facts while preserving provenance and evidence.

\stitle{Extraction as a physical operator.}
Extraction should not be an offline ETL script placed before the database. It should be a first-class physical operator whose input is a materialization gap and whose output is typed tuples, provenance, confidence, coverage updates, and failures~\cite{datamosaic2026}. This operator must support projection pushdown, predicate pushdown, batching, entity resolution, validation, and incremental maintenance. Treating extraction as an operator gives the optimizer something concrete to cost, reorder, cache, and combine with relational operators, while making
missing evidence, conflicting evidence, and low-confidence values explicit outputs rather than hidden prompt artifacts.

\stitle{Transactions over data and evolution.}
Transaction management must be revisited. In \sys, many writes may run concurrently: universal-relation changes add roles, schema-evolution actions add attributes or relationship tables, extraction jobs add tuples, validation jobs repair values, and user queries may read partially materialized state. The system needs isolation levels for evolving databases: should a query see freshly extracted tuples before validation, should related universal-relation, schema, and tuple deltas commit atomically, and should failed extraction calls roll back or leave partial evidence?
Without such semantics, the same natural-language workload may produce non-repeatable answers, inconsistent coverage, and irreproducible schema states.

\stitle{Semantic operators for multimodal evidence.}
\sys needs semantic operators that treat interpretation itself as part of query processing. Document collections mix text, tables, forms, figures, screenshots, and images; a query may require reading a clause, locating a table cell, comparing a chart, or linking a visual region to an entity. These operations should become typed operators with inputs, outputs, costs, confidence, provenance, and failure modes, so the optimizer can combine them with relational operators and the system can materialize their results as reusable state.
%!TEX root = ../main.tex
\section{Conclusion}
\label{sec:call-to-arms}

RAG made documents searchable; \sys asks the database question: what durable, governed state should each query leave behind? The answer is not a schema-first replica of the corpus, but a query-shaped database whose universal relation, logical schema, materialized facts, coverage, indexes, and constraints evolve together while preserving logical path independence for users. The agenda is clear: make evolution---universal-relation deltas, schema deltas, extraction deltas, coverage deltas, validation, indexing, and transactions---a first-class DBMS operation.

\bibliographystyle{ACM-Reference-Format}
\bibliography{references}

\end{document}